**Electrical Control of Linear Dichroism in Black Phosphorus from the Visible to Mid-Infrared**

Michelle C. Sherrott[1,2,‡], William S. Whitney[3,‡], Deep Jariwala[1,2], Cora M. Went[2,3], Joeson Wong[1], George R. Rossman[4], Harry A. Atwater[1,2,5]

1. Thomas J. Watson Laboratory of Applied Physics, California Institute of Technology, Pasadena, CA 91125, USA
2. Resnick Sustainability Institute, California Institute of Technology, Pasadena, CA 91125, USA
3. Department of Physics, California Institute of Technology, Pasadena, CA 91125, USA
4. Division of Geological and Planetary Sciences, California Institute of Technology, Pasadena, CA 91125, USA
5. Joint Center for Artificial Photosynthesis, California Institute of Technology, Pasadena, CA 91125, USA

[‡] Equal contributors

*Corresponding author: Harry A. Atwater (haa@caltech.edu)

**Abstract**

*The incorporation of electrically tunable materials into photonic structures such as waveguides and metasurfaces enables dynamic control of light propagation by an applied potential. While many materials have been shown to exhibit electrically tunable permittivity and dispersion, including transparent conducting oxides (TCOs) and III-V semiconductors and quantum wells, these materials are all optically isotropic in the propagation plane. In this work, we report the first known example of electrically tunable linear dichroism, observed here in few-layer black phosphorus (BP), which is a promising candidate for multi-functional, broadband, tunable photonic elements. We measure active modulation of the linear dichroism from the mid-infrared to visible frequency range, which is driven by anisotropic quantum-confined Stark and Burstein-Moss effects, and field-induced forbidden-to-allowed optical transitions. Moreover, we observe high BP absorption modulation strengths, approaching unity for certain thicknesses and photon energies.*

*Introduction:*

As photonic structures for controlling the near- and far-field propagation of light become increasingly complex and compact, the need for new materials that can exhibit unique, strong light-matter interactions in the ultra-thin limit is growing rapidly. Ultrathin van der Waals materials are especially promising for such applications, as they allow for the control of light at the atomic scale, and have properties that can be modulated actively using an external gate voltage[1,2]. Of these, few-layer black phosphorus (BP) is particularly noteworthy due to its high electronic mobility, and a direct band gap that can be tuned as a function of thickness from 0.3 eV to 2 eV[3,4]. This has enabled the realization of numerous optoelectronic devices with high performance, including photodetectors that can be easily integrated with other photonic elements such as waveguides[5–9]. In addition to this static control, recent works using electrostatic gating and potassium ions have shown that the electronic band gap of BP may be tuned by an electric field.[10–12]

One of the most salient features of BP is its large in-plane structural anisotropy, leading to a polarization-dependent optical response[13,14] as well as mechanical[15], thermal[16], and electrical transport characteristics[17,18] that vary with in-plane crystallographic orientation[19]. This optical anisotropy corresponds to a large, broadband birefringence[20], wherein the distinct optical index of refraction along each axis leads to a phase delay between polarization states of light. Moreover, mirror-symmetry in the x-z plane forbids intersubband optical transitions along the zigzag axis, and as a result, BP exhibits significant linear dichroism, wherein the material absorption depends strongly on the polarization state of exciting light[14,21].

In this work, we experimentally demonstrate that the application of a static electric field enables the modulation of the linear dichroism of few-layer black phosphorus (BP). This response – which approaches near-unity modulation of the BP oscillator strength for some thicknesses and photon energies – is achieved by active control of quantum-confined Stark and Burstein-Moss effects, and of quantum-well selection rules. We observe anisotropic modulation from the visible to mid-infrared (mid-IR) spectral regimes, behavior not seen in traditional electro-optic materials such as graphene[22], transparent conducting oxides[23,24], silicon[25], and quantum wells[26]. This opens up the possibility of realizing novel photonic structures in which linear dichroism in the van der Waals plane can be continuously tuned with low power consumption, because the switching is electrostatic in nature. By controlling optical losses in the propagation plane, for example, efficient in-plane beam steering of surface plasmon polaritons or other guided modes is enabled. Moreover, a tunable polarizer could be realized by the tuning of the polarization state of light absorbed in a resonant structure containing BP. Because this modulation is strongest at infrared wavelengths, it could also enable control of the polarization state of thermal radiation[27–29].

## *Results*

**Infrared Tunable Linear Dichroism in a Thin BP Flake**

In order to probe and distinguish the electro-optical tuning mechanisms evident in few-layer BP, we used a combination of gating schemes wherein the BP either floats in an applied field or is contacted, as shown in Fig. 1a and described further in the Methods section. Polarization-dependent optical measurements are taken aligned to the crystal axes, in order to probe the structural anisotropy shown in Fig. 1b. This enables us to isolate the contribution of charge-carrier density effects – i.e. a Burstein-Moss shift – and external field-effects – i.e.: the quantum-confined Stark effect and control of forbidden transitions in the infrared – to the modulation of linear dichroism, qualitatively illustrated in Figures 1c and 1d[30–33]. In the anisotropic Burstein-Moss (BM) shift, the optical band gap of the material is changed as a result of band filling and the consequent Pauli-blocking of intersubband transitions. As the carrier concentration of the sample is changed, the Fermi level moves into (out of) the conduction or valence band, resulting in a decrease (increase) of absorptivity due to the disallowing (allowing) of optical transitions[34,35]. Because intersubband optical transitions are only allowed along the armchair axis of BP, this modulation occurs only for light polarized along this axis. In the quantum-confined Stark Effect, the presence of a strong electric field results in the leaking of electron and hole wave functions into the band gap as Airy functions, red-shifting the intersubband transitions energies[36]. In quantum well structures, this red-shifting is manifested for multiple subbands, and therefore can be observed over a wide range of energies above the band gap. To assess the gate-tunable anisotropy of the optical response of BP, the armchair and zigzag axes, illustrated in Fig. 1e, of the samples considered are identified by a combination of cross-polarized visible microscopy, described in the Supplementary Information, and either polarization-dependent Raman spectroscopy or infrared measurements, described below. Representative Raman spectra are presented for the visible frequency sample on SrTiO$_3$ in Figure 1f. The optically active armchair axis exhibits a maximum intensity of the $A_g^2$ resonant shift at 465 cm$^{-1}$, whereas this is a minimum for the zigzag axis[37].

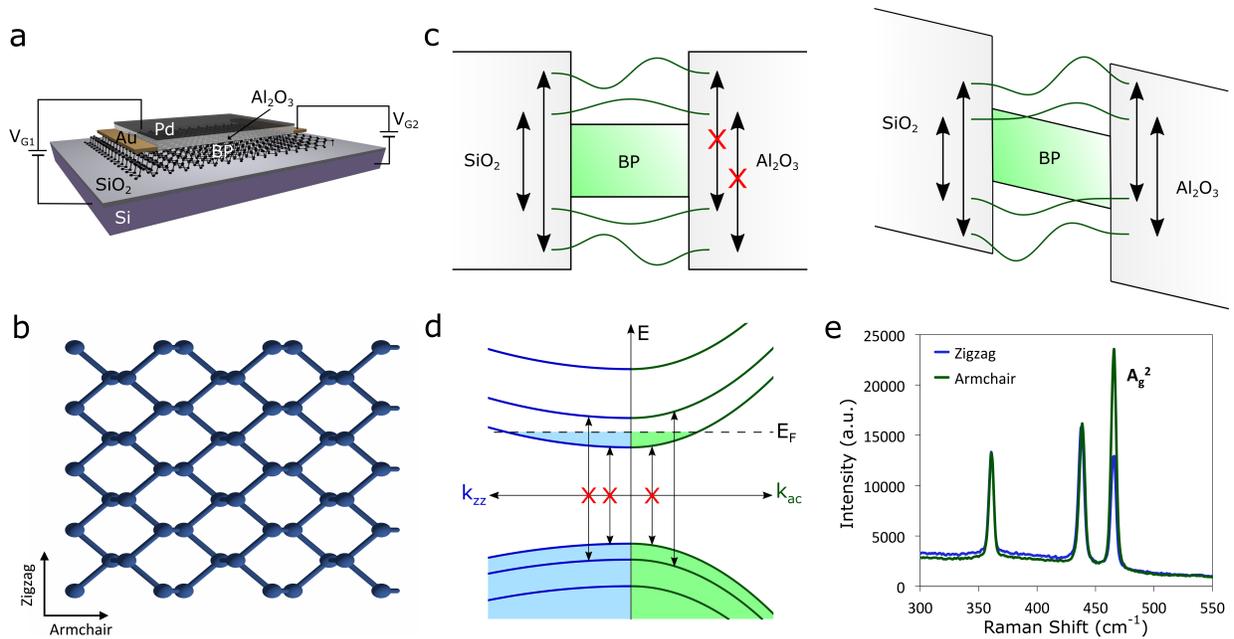

*Figure 1 | Anisotropic electro-optical effects in few-layer BP. (a),* Schematic figure of infrared modulation devices. Few-layer BP is mechanically exfoliated on 285 nm SiO$_2$/Si and then capped with 45 nm Al$_2$O$_3$ by ALD. A semitransparent top contact of 5 nm Pd is used to apply field ($V_{G1}$) while the device floats and 20 nm Ni/200 nm Au contacts are used to gate ($V_{G2}$) the contacted device. *(b),* Crystal structure of BP with armchair and zigzag axes indicated. *(c),* Illustration of quantum-confined Stark effect and symmetry-breaking effect of external field. Under zero external field, only optical transitions of equal quantum number are allowed. An external field tilts the quantum well-like energy levels, causing a red-shifting of the optical band gap and allowing previously forbidden transitions. *(d),* Illustration of anisotropic Burstein-Moss shift in BP. Intersubband transitions are blocked due to the filling of the conduction band. Along the ZZ axis, all optical transitions are disallowed regardless of carrier concentration. *(e),* Raman spectra with excitation laser polarized along AC and ZZ axes. The strength of the $A_g^2$ peak is used to identify crystal axes.

To illustrate the mechanisms of tunable dichroism of BP in the mid-infrared, we measure modulation of transmittance using Fourier-Transform Infrared (FTIR) microscopy as a function of externally ($V_{G1}$) or directly applied bias ($V_{G2}$), presented for a 3.5 nm thick flake, as determined from atomic force microscopy (AFM) presented in Supplementary Fig. 1, in Figure 2. Fig. 2a presents the raw extinction of the flake along the armchair axis at zero bias, obtained by normalizing the armchair axis to the optically inactive zigzag axis. A band edge of approximately 0.53 eV is measured, consistent with a thickness of 3.5nm. A broad, weak shoulder feature is observed at approximately 0.75 eV. The corresponding calculated optical constants for the flake are presented in Figure 2c for comparison.

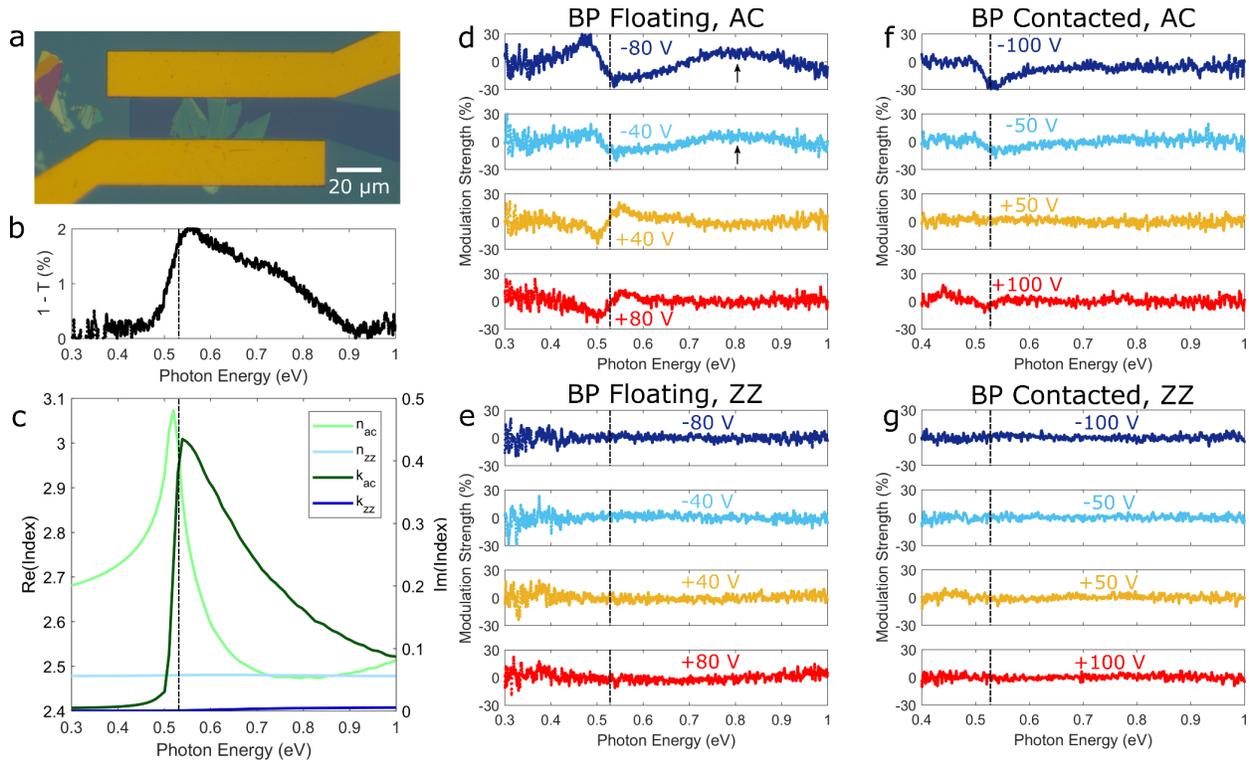

*Figure 2 | Electrically tunable linear dichroism: quantum-confined Stark and Burstein-Moss effects and forbidden transitions. (a),* Optical image of fabricated sample. *(b),* Zero-bias infrared extinction of 3.5 nm flake, polarized along armchair (AC) axis. *(c),* Calculated index of refraction for 3.5 nm thick BP with a Fermi energy at mid-gap. *(d),* Modulation of BP oscillator strength with field applied to floating device, for light polarized along the AC axis. *(e),* Corresponding modulation for light polarized along the zigzag (ZZ) axis. *(f),* Modulation of BP oscillator strength with gating of contacted device, for light polarized along the AC axis. *(g),* Corresponding modulation for light polarized along the ZZ axis.

Figures 2d and 2e illustrate the influence of an external field on the extinction of BP with carrier concentration held constant (i.e. the BP is left floating). The extinction data for each voltage is normalized to the zero bias case and to the peak BP extinction seen in Figure 2b, to obtain a modulation strength percentage that quantifies the observed modulation of the BP oscillator strength. We note that this normalization scheme underestimates modulation strength away from the band edge, where BP extinction is maximal. Along the armchair axis, presented in Fig. 2d, two modulation features are measured near photon energies of 0.5 and 0.8 eV. We explain the first feature at 0.5 eV as arising from a shifting of the BP band edge due to the quantum-confined Stark effect. At negative bias, the band gap effectively shrinks and this is manifest as a redistribution of oscillator strength near the band edge to lower energies. As a result, an increase in absorptance is measured below the zero-bias optical band gap, and a decrease is seen above it. At positive bias, this trend is weakened and reversed. We propose two explanations for this asymmetry: the first is the influence of electrical hysteresis, and the second is the presence of a small internal field in the BP at zero bias, which has been observed in previous works on the infrared optical response of few-layer BP[14].

The second, higher energy feature observed in the measured spectrum does not correspond to any predicted intersubband transition. Rather, we propose it arises due to the allowing of an optical transition that was previously forbidden by quantum-well selection rule constraints dictated by symmetry (i.e. only transitions of equal quantum number are allowed under zero field[31]). We note that this feature is present in the 0 V extinction spectrum, consistent with a zero-bias internal field. As the symmetry is further broken with an externally-applied electric field, this transition is strengthened. Under positive bias, the internal and external fields are in competition, resulting in minimal change. This suppressed modulation can also be attributed to hysteresis, as before.

In Figure 2e, no modulation is measured for any applied bias for light polarized along the zigzag axis. This can be well understood due to the dependence of the Stark effect on the initial oscillator strength of an optical transition; because no intersubband optical transitions are allowed along this axis, the field effect is weak. Similar behavior has been observed in excitons in ReS$_2$ based on an optical Stark effect[38]. Moreover, while the externally applied field can allow 'forbidden' transitions along the armchair axis by breaking the out-of-plane symmetry of the quantum well, in-plane symmetry properties and thus the selection rule precluding all zig-zag axis intersubband transitions are unaffected. This selection rule and the corresponding symmetry properties have been previously described [17].

In Figures 2f and 2g, we present the complementary data set of tunable dichroism measurements due to a directly applied gate bias with electrical contact made to the BP in a standard field-effect transistor (FET) geometry. Here, we observe modulation dominated by carrier concentration effects. At the band gap energy of approximately 0.53 eV, a simple decrease in absorptance is observed at negative and large positive biases, consistent with an ambipolar BM shift. Unlike the results of applying field while the BP floats, no modulation of the forbidden transition at 0.75 eV is observed; this is explained in part due to the screening of the electric field due to the carrier concentration modulation. We additionally may consider the possibility that this optical transition is disallowed by Pauli-blocking effects, negating the symmetry-breaking effect of the directly applied field. As in the case for the floating BP measurement, no modulation is observed along the zigzag axis.

**Infrared Tunable Linear Dichroism in a Thick BP Flake**

The anisotropic electro-optical effects described above change character rapidly as the BP thickness – and hence band gap and band structure – is varied. Figure 3 presents analogous results on a flake of 8.5 nm thickness, determined by AFM (see Supplementary Fig. 1), for which an optical image is presented in Fig. 3a. Due to the increased thickness, the energy separation between subbands is smaller, resulting in a narrower free-spectral range between absorptance features measured in the zero-bias spectrum, presented in Fig. 3b and for which corresponding calculated optical constants are presented in Fig. 3c. Results for modulation by an external field with the BP left floating are presented in Fig. 3d. As in the thin flake, substantial modulation of the absorptance at each intersubband transition is observed due to the QCSE red-shifting the energy of the subbands. Due to the large Stark coefficient in BP – which increases with thickness in the few-layer limit – absorption is nearly 100% suppressed, resulting in an approximately isotropic optical response from the material[11,39]. Unlike the previous sample, modulation of forbidden transitions is not apparent; all features correspond to transitions measured in the 0 V normalization scheme as well as the calculated optical constants for a thickness of 8.5 nm. As before, no modulation is seen along the zig-zag axis, as shown in Supplementary Fig. 2. In Fig. 3e, the modulation for directly gated, contacted BP is shown. The observed modulation – a reduction in extinction centered at each of the calculated intersubband transition energies – is relatively weak and does not persist to high photon energies. This suggests that the dominant modulation mechanism is the ambipolar BM shift, rather than the QCSE. Additional measurements at lower energies are presented in Supplementary Fig. 3.

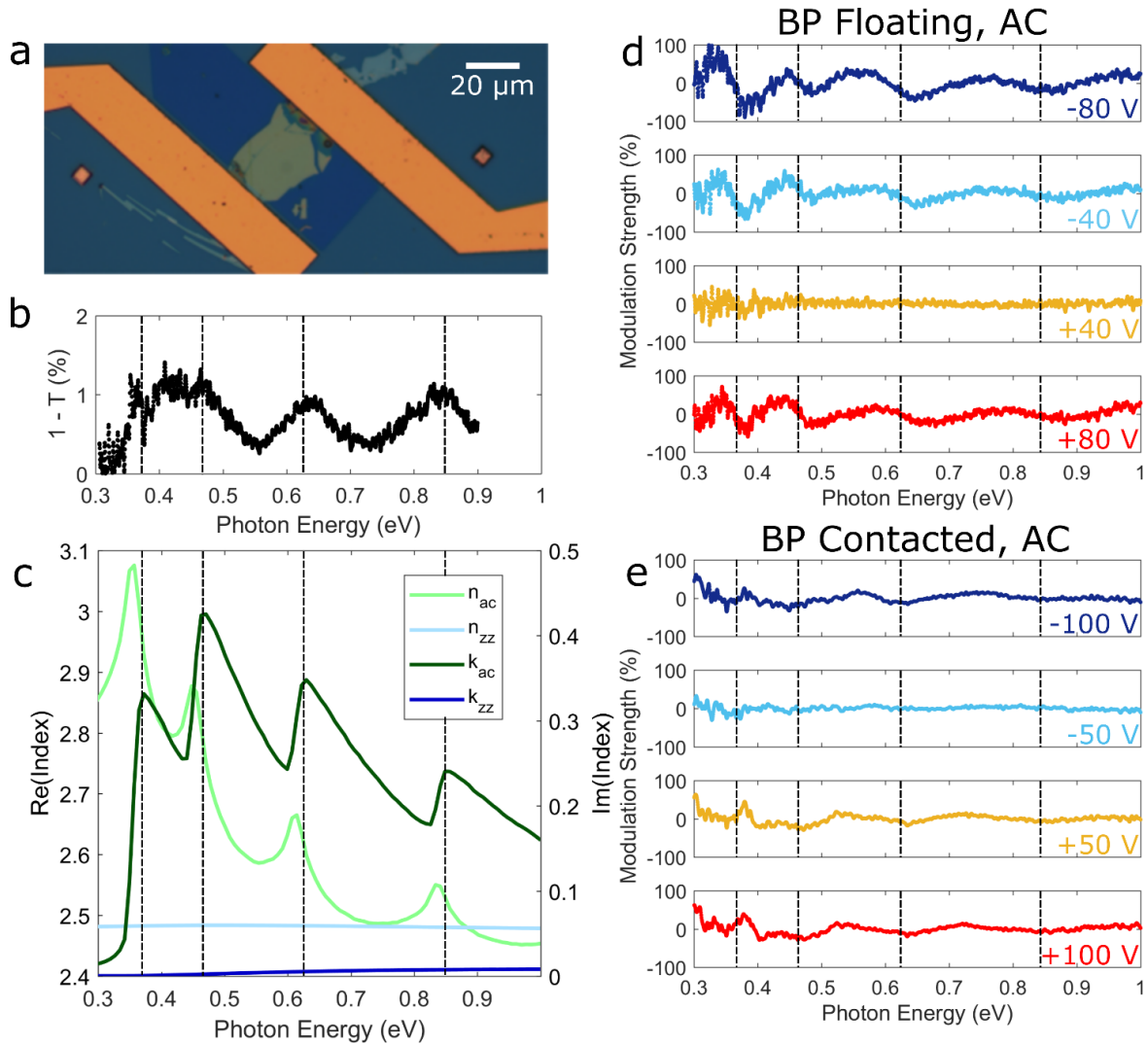

*Figure 3 | Variation of Modulation with BP Thickness. (a),* Optical image of fabricated 8.5 nm sample. *(b),* Zero-bias extinction of 8.5 nm flake, polarized along AC axis. *(c),* Calculated index of refraction for 8.5 nm thick BP. *(d),* Modulation of BP oscillator strength with field applied to floating device, for light polarized along the AC axis. *e,* Modulation of BP oscillator strength with gating of contacted device, for light polarized along the AC axis.

**Visible-to-Near Infrared Tunable Linear Dichroism in a Thick BP Flake**

Finally, in Figure 4 we present results of gate-tunable dichroism at visible frequencies in a 20 nm thick flake, comparable to those considered for infrared modulation. A new device geometry is used to enable transmission of visible light, shown schematically in Fig. 4a and in an optical image in Fig. 4b. In this configuration, a SrTiO$_3$ substrate is utilized to allow transmission-mode measurements at visible wavelengths. A symmetric gating scheme is devised based on semi-transparent top and back gate electrodes of 5 nm Ni, as described in the Methods section. Only an applied field, floating BP measurement is utilized, as band-filling effects should be negligible at this energy range. In Fig. 4c, we present modulation results from 1.3 to 2 eV. Due to the QCSE, modulation is observed up to 1.8 eV, corresponding to red light. Thus we demonstrate that electro-optic modulation of linear dichroism is possible across an extraordinarily wide range of wavelengths in a single material system, enabling multifunctional photonic devices with broadband operation.

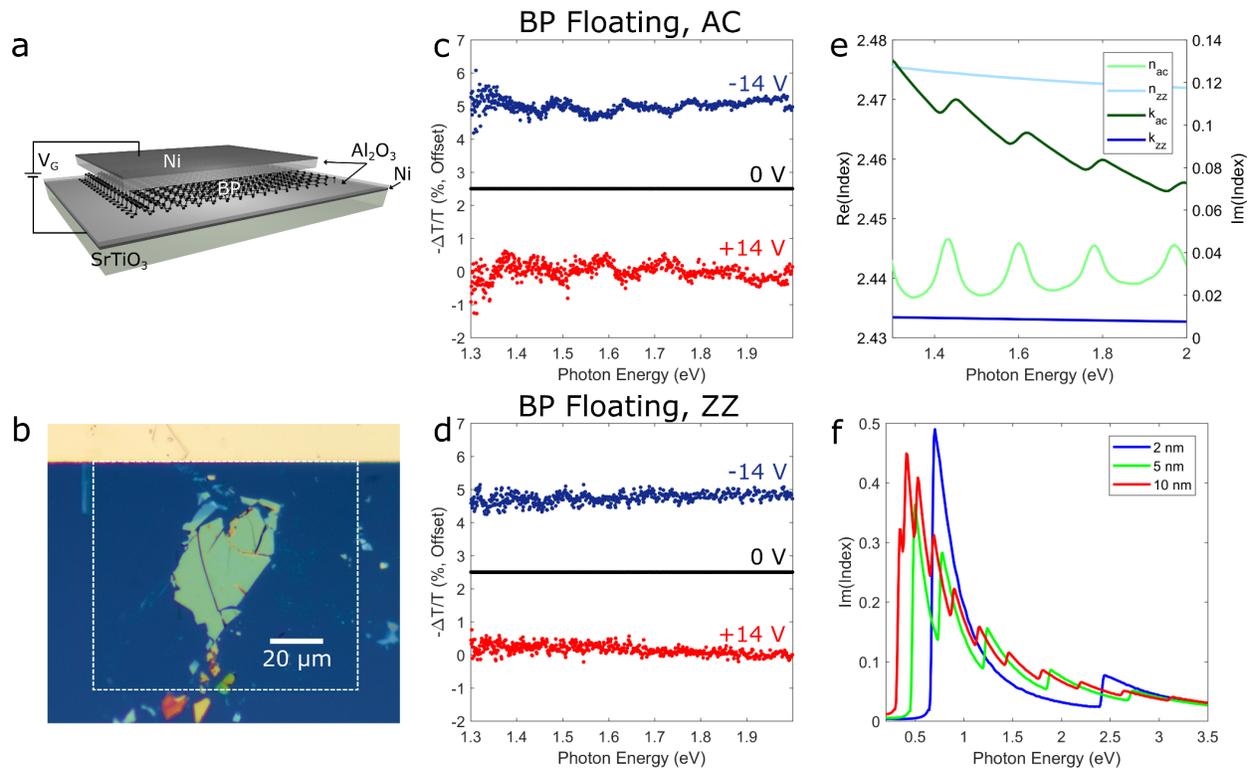

*Figure 4 | Modulation in the Visible. (a),* Schematic figure of visible modulation device. Few-layer BP is mechanically exfoliated on 45 nm $Al_2O_3$/5 nm Ni on $SrTiO_3$ and then coated with 45 nm $Al_2O_3$. A 5 nm thick semitransparent Ni top contact is used. *(b),* Optical image of fabricated sample with 20 nm thick BP. Dashed white line indicates the boundary of the top Ni contact. *(c),* Modulation of extinction with field applied to floating device, for light polarized along the AC axis. *(d),* Corresponding modulation for light polarized along the ZZ axis. *(e),* Calculated index of refraction for 20 nm thick BP for the measured energies. *(f),* Calculated imaginary index of refraction of several thicknesses of BP from the infrared to visible.

*Discussion*

The decay of BP intersubband oscillator strength at higher photon energies provides a spectral cutoff for QCSE-based modulation, but for 5 nm BP or thinner this oscillator strength is strong through the entire visible regime, as illustrated in Fig. 4f. We thus suggest that in very thin BP, strong modulation of absorption and dichroism is possible to even higher energies. By selecting a flake of 2 nm, for example, tunable linear dichroism is possible up to 3 eV from the band gap energy of 0.75 eV. A higher density of features, beginning at lower energies, may be introduced by utilizing a thicker flake, with slightly decreased modulation strength, as seen for 5 and 10 nm thickness flakes. We also note that by substituting graphene top and bottom contacts or utilizing nanophotonic techniques to focus light in the BP, higher absolute modulation strength could be easily realized.

This phenomenon is in stark contrast to the gate-tunability of the optical response of other 2D materials, where substantial modulation is typically constrained to the narrowband energy of the primary exciton, as in $MoS_2$ and $WS_2$[1,40]. In another van der Waals materials system, monolayer graphene, tunability is accessible over a broader wavelength range due to the Pauli-blocking of optical transitions at $2E_F$; however, this is limited to the range over which electrostatic gating is effective, typically between $E_F \sim 0$ to $E_F \sim 0.5$ eV[2,41]. Moreover, these materials are not dichroic or birefringent in-plane, and so BP offers a novel phenomenon that can be taken advantage of to realize previously challenging or impossible photonic devices. The same restriction is true of bulk tunable materials such as quantum wells, transparent conducting oxides, and transition metal nitrides.

In summary, we have demonstrated broadly tunable linear dichroism in few-layer black phosphorus. We can explain this modulation as arising from a combination of quantum-confined Stark effects, ambipolar Burstein-Moss effects, and the allowing of forbidden optical transitions by the symmetry-breaking effects of the applied electric field. We identify the different physical mechanisms governing this tunability by comparing the modulation response from a dual gate wherein the BP is left floating to a single gate directly applied to the BP, leading to modulation of carrier concentration.  By varying the thickness, and therefore band structure of the BP, we see that it is possible to control the spectroscopic modulation as well as the dominant physical mechanisms of modulation. We suggest that this phenomenon is a promising platform for controlling the in-plane propagation of surface or waveguide modes, as well as for polarization-switching, reconfigurable far-field metasurfaces. These applications are particularly promising in light of our observation that BP absorption can be modulated from anisotropic to nearly isotropic in-plane.  Because van der Waals materials can be easily integrated into photonic devices, this promises to introduce new functionalities that cannot be realized by conventional electro-optic materials.

## Methods

**Infrared Sample Preparation:**

Samples for infrared measurements were fabricated by mechanically exfoliating few-layer BP onto 285 nm $SiO_2$/Si in a glove box environment. Contacts of 20 nm Ni/200 nm Au were fabricated by electron beam lithography, electron beam evaporation, and liftoff. A top gate dielectric of 45 nm $Al_2O_3$ was deposited by atomic layer deposition (ALD) following the technique in ref[42], and a semi-transparent top contact of 5 nm Ni was deposited by electron beam evaporation and liftoff. Measurements were performed in a Fourier Transform Infrared Spectrometer coupled to a microscope.

**Visible Sample Preparation:**

Samples for visible measurements were fabricated by depositing a 5 nm thick semi-transparent back contact of Ni, followed by 45 nm $Al_2O_3$ by ALD on a 0.5 mm thick $SrTiO_3$ substrate. Few-layer BP was then mechanically exfoliated and electrical contacts were fabricated as above. Measurements are performed in a visible spectrometer. Nickel was selected as the optimum metallic contact through Finite-Difference Time Domain simulations presented in Supplementary Fig. 4.

**Calculations:**

Calculations of the optical constants of BP are based on the formalism developed in ref[31]. Optical conductivity $\sigma$ is calculated using the Kubo formula within an effective low-energy Hamiltonian for different thicknesses. The permittivity is calculated as $\varepsilon(\omega) = \varepsilon_\infty + i\sigma/\omega\delta$ where $\delta$ is the thickness of the BP, and the high-frequency permittivity $\varepsilon_\infty$ is taken from ref[43].

## Acknowledgements:

The authors gratefully acknowledge support from the Department of Energy, Office of Science under Grant DE-FG02- 07ER46405 and for facilities of the DOE "Light-Material Interactions in Energy Conversion" Energy Frontier Research Center (DE-SC0001293). W.S.W. also acknowledges support from an NDSEG Graduate Research Fellowship. M.C.S., D.J. and C.M.W. acknowledge fellowship support from the Resnick Institute. J.W. acknowledges support from the National Science Foundation Graduate Research Fellowship under grant no. 1144469.

## Author Contributions:

M.C.S, W.S.W., D.J., and H.A.A. conceived the experiment with support from G.R.R. Samples were prepared by M.C.S, W.S.W. and D.J. and measured by W.S.W. and M.C.S. C.M.W. performed Raman measurements. J.W. assisted in figure preparation. Theoretical works on the index of refraction of BP were done by W.S.W. Full-wave simulations were done by M.C.S. Interpretation of results and preparation of the manuscript was done with contributions from all authors.